\documentclass[12pt]{article}

\usepackage{graphicx}
\usepackage{amsmath}
\usepackage{amssymb}
\allowdisplaybreaks

\begin{document}

\baselineskip=17.5pt plus 0.2pt minus 0.1pt

\renewcommand{\theequation}{\arabic{equation}}
\renewcommand{\thefootnote}{\fnsymbol{footnote}}
\makeatletter
\def\CR{\nonumber \\}
\def\pt{\partial}
\def\be{\begin{equation}}
\def\ee{\end{equation}}
\def\bea{\begin{eqnarray}}
\def\eea{\end{eqnarray}}
\def\eq#1{(\ref{#1})}
\def\la{\langle}
\def\ra{\rangle}
\def\hyp{\hbox{-}}
\def\sixj#1#2#3#4#5#6{\left\{ \begin{matrix} #1&#2&#3 \\ #4&#5&#6 \end{matrix} \right\}}
\def\threej#1#2#3#4#5#6{\left( \begin{matrix} #1&#2&#3 \\ #4&#5&#6 \end{matrix} \right)}

\begin{titlepage}
\title{\hfill\parbox{4cm}{ \normalsize YITP-06-27 \\{\tt hep-th/0606066}}\\
\vspace{1cm} Tensor model and dynamical generation of \\ commutative nonassociative fuzzy spaces}
\author{
Naoki {\sc Sasakura}\thanks{\tt sasakura@yukawa.kyoto-u.ac.jp}
\\[15pt]
{\it Yukawa Institute for Theoretical Physics, Kyoto University,}\\
{\it Kyoto 606-8502, Japan}}
\date{\normalsize July, 2006}
\maketitle
\thispagestyle{empty}

\begin{abstract}
\normalsize
Rank-three tensor model may be regarded as theory of dynamical fuzzy spaces, because a fuzzy space is 
defined by a three-index coefficient of the product between functions on it, $f_a*f_b={C_{ab}}^cf_c$.  
In this paper, this previous proposal is applied to dynamical generation of commutative nonassociative fuzzy spaces.
It is numerically shown that fuzzy flat torus and fuzzy spheres of various dimensions 
are classical solutions of the rank-three tensor model. 
Since these solutions are obtained for the same coupling constants of the tensor model, 
the cosmological constant and the dimensions are not fundamental but can be regarded as dynamical quantities. 
The symmetry of the model under the general linear transformation can be identified with 
a fuzzy analog of the general coordinate transformation symmetry in general relativity. 
This symmetry of the tensor model is broken at the classical solutions.
This feature may make the model to be a concrete finite setting for applying the old idea of obtaining gravity 
as Nambu-Goldstone fields of the spontaneous breaking of the local translational symmetry. 

\end{abstract}
\end{titlepage}

\section{Introduction}
The tensor model is an interesting generalization of the matrix model. While the matrix model gives an 
analytical method to describe the two-dimensional simplicial quantum gravity \cite{Ambjorn:1997di}, 
the tensor model was proposed to describe the simplicial quantum gravity in higher dimensions 
and dynamical lattice topological  field theories \cite{Ambjorn:1990ge,Sasakura:1990fs,Godfrey:1990dt,Boulatov:1992vp, Ooguri:1992eb,DePietri:1999bx,DePietri:2000ii, Freidel:2005cg}.
Feynman diagrams of the models are assumed to correspond to dual diagrams of simplicial complexes. 
In the matrix model, the topological expansion of the two-dimensional manifolds can be realized in 
the large $N$ expansion, and this fact is essentially used in the computation.
Unfortunately, this kind of expansion has not been realized in the tensor model so far, and it seems hard
to use the tensor model as an analytical tool to study quantum gravity.

Another interpretation of rank-three tensor model was proposed by the present
author in \cite{Sasakura:2005js}. A fuzzy space can be 
characterized by a rank-three tensor ${C_{ab}}^c$, which defines the algebraic relation $f_a* f_b={C_{ab}}^c f_c$ 
between functions $f_a$ on a fuzzy space. Therefore dynamical fuzzy spaces may be described by
theory which contains the rank-three tensor as a dynamical variable. A fuzzy space will be obtained as 
a classical solution to the equation of motion of a tensor model. Unlike the original interpretation
in the preceding paragraph, it will be easier to deduce physical results from the tensor model
under this new interpretation.

An important advantage of the tensor model as theory of spacetime 
is the existence of the symmetry under a fuzzy analog of the general coordinate transformation
in general relativity. 
As discussed in \cite{Sasakura:2005js}, the general linear transformation on the rank-three tensor 
can be naturally identified with it. 
Around the classical solutions, the symmetry will be broken and will be non-linearly realized,
which is actually the same situation for the general coordinate transformation on a geometric background.
There is an old idea 
that gravity may be obtained as the Nambu-Goldstone fields of the spontaneous breaking of the local 
translational symmetry \cite{Borisov:1974bn,Boulanger:2006tg}.
Therefore the tensor model may provide a concrete finite model for the old idea.
 
The notion of noncommutative coordinates \cite{snyder}-\cite{connes} is a natural application of 
the quantum procedure to spacetime. 
In fact, noncommutative field theories are derived as effective field theories on D-branes in string theory
with a background $B_{\mu\nu}$ field 
\cite{Seiberg:1999vs} and of three-dimensional quantum gravity coupled with matter \cite{Freidel:2005bb}.
On the other hand, nonassociativity is also known to appear in open string theory with a non-constant background 
$B_{\mu\nu}$ field \cite{Cornalba:2001sm}. It was also argued that 
the algebra of closed string field theory should be commutative nonassociative \cite{Witten:1985cc}. 
Therefore, though its relation to quantization of spacetime is not clear, 
it would be fair to say that nonassociativity is also another physically sensible structure of spacetime.
Moreover, it is generally much easier to obtain commutative nonassociative fuzzy spaces of physical interest 
\cite{Ramgoolam:2001zx, deMedeiros:2004wb} than noncommutative ones which seem to
generally require a kind of symplectic structure. 
In addition, quantum field theory on commutative 
nonassociative spacetimes seems to be able to respect the principles in physics more faithfully \cite{Sasai:2006ua}
than noncommutative quantum field theory, which is known to have
some unusual properties such as the UV/IR mixing \cite{Filk:1996dm,Minwalla:1999px} 
and the violation of causality \cite{Seiberg:2000gc} and unitarity \cite{Gomis:2000zz}.

The main purpose of this paper is to show the existence of some elementary commutative nonassociative fuzzy 
spaces as classical solutions of the rank-three tensor model. 

This paper is organized as follows. In the following section, the rank-three tensor model is defined. A constraint is 
imposed on its dynamical variable to simplify the original model in \cite{Sasakura:2005js}. 
This constraint corresponds to dealing with commutative nonassociative fuzzy spaces. 
In Section \ref{simpleexample}, a simple example of a commutative 
nonassociative space is given. This space is not a solution to the tensor model, but provides a good reference to
understand the qualitative features of the numerical solutions.
In Section \ref{numericalsolutions}, the numerical solutions representing
fuzzy flat torus and fuzzy spheres of various dimensions are obtained and checked.   
The final section is devoted to summary and discussions.
  
\section{The model}
\label{themodel}
The original model proposed in \cite{Sasakura:2005js} contains  
a symmetric tensor $g^{ab}$ and a rank-three tensor $C_{abc}$
as real dynamical variables.
Without any constraints on the variable $C_{abc}$, there are a lot of possibilities of constructing actions, 
and the analysis to find solutions to equation of motion becomes quite complicated. 
To simplify the model\footnote{Since the discussions in this paper are on classical solutions, the constraint
may be regarded as being imposed on classical solutions rather than on the variable.}, 
a constraint will be imposed on the variable $C_{abc}$. 
To allow fuzzy spaces to become good approximations to ordinary spaces, 
I assume such a constraint be also satisfied by ordinary spaces. 

The algebra of functions on an ordinary continuum space is commutative and associative. 
Let $\{f_a\}$ denote a complete set of independent functions on such a space.
A rank-three tensor defines the product between two functions,
\be
f_a * f_b={C_{ab}}^c f_c.
\ee
The commutativity implies
\be
\label{commutativity}
{C_{ab}}^c={C_{ba}}^c,
\ee
while the associativity implies
\be
\label{associativity}
{C_{ab}}^d {C_{dc}}^e={C_{ad}}^e {C_{bc}}^d.
\ee
Let me define a tensor
\be
\eta_{ab} \equiv {C_{ac}}^d{C_{bd}}^c,
\ee
which is obviously symmetric for the indices.
Then let me consider 
\be
C_{abc}\equiv {C_{ab}}^d \eta_{dc}.
\ee
By using the associativity \eq{associativity}, the right-hand side can be rewritten, and
\be
C_{abc}={C_{ad}}^f {C_{be}}^d {C_{cf}}^e.
\ee 
This shows that the tensor $C_{abc}$ is symmetric under the cyclic permutations of the indices.
Therefore, from the commutativity \eq{commutativity}, 
$C_{abc}$ is totally symmetric under the permutations of the indices.
If $\eta_{ab}$ is assumed to be not singular, $g^{ab}$ can be identified with $(\eta_{ab})^{-1}$. 
Thus the constraint to be imposed on $C_{abc}$ is the total symmetry for the indices, 
\be
\label{constraintc}
C_{abc}=C_{bca}=C_{cab}=C_{bac}=C_{acb}=C_{cba}.
\ee
Conversely, if one starts with $C_{abc}$ with the constraint \eq{constraintc},
the constraint is not enough to show the associativity \eq{associativity},
and the algebra generated by ${C_{ab}}^c=C_{abc'}g^{cc'}$ will be a commutative nonassociative (or
associative) algebra in general.

\begin{figure}
\begin{center}
\includegraphics[scale=.7]{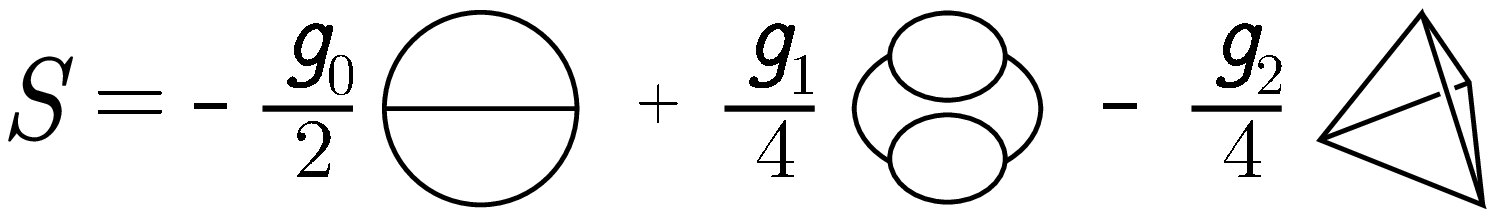}
\caption{The graphical representation of the action. A three-vertex represents $C_{abc}$ and a line $g^{ab}$.}
\label{fig:s012}
\end{center}
\end{figure}

Under the constraint \eq{constraintc}, the variety of action is drastically simplified, compared with the original
proposal \cite{Sasakura:2005js}. 
In the quadratic order of $C_{abc}$, there is only one possibility,
\be
S_0=C_{abc}C^{abc}.
\ee
In the quartic order, there are only two possibilities,
\bea
S_1&=&C_{abc}C^{dbc}C_{def}C^{aef}, \\
S_2&=&C_{abc}{C^{ad}}_e {C^b}_{df}C^{cef}.
\eea
The total action is given with three coupling constants $g_i$ by
\be
\label{totals}
S=-\frac{g_0}{2} S_0+\frac{g_1}{4}S_1 -\frac{g_2}4 S_2,
\ee 
which is graphically shown in Fig.\ref{fig:s012}.

This action is invariant under the $GL(n,R)$ transformation,
\bea
\label{ctrans}
C'_{abc}&=&{M_a}^{a'} {M_b}^{b'}{M_c}^{c'}\,C_{a'b'c'}, \cr
g'^{ab}&=& {{M^{-1}}_{a'}}^a{{M^{-1}}_{b'}}^b\, g^{a'b'},
\eea
where $n$ is the total number of functions on a fuzzy space and $M\in GL(n,R)$. 
As discussed in \cite{Sasakura:2005js}, 
the transformation can be identified with a fuzzy analog of the general coordinate transformation. 

\begin{figure}
\begin{center}
\includegraphics[scale=.8]{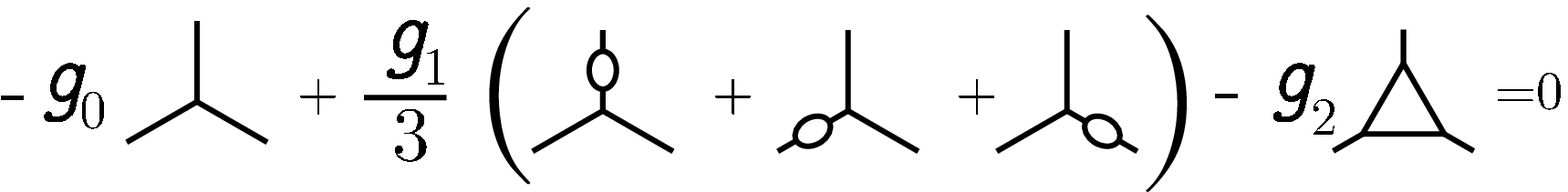}
\caption{The graphical representation of the equation of motion.}
\label{fig:eom}
\end{center}
\end{figure}

The equation of motion for $C_{abc}$ from $S$ in \eq{totals} is given by
\be
\label{eom}
-g_0\, C_{abc}+\frac{g_1}3\,\left[ C_{ade}C^{a'de}C_{a'bc}+({\rm cyclic\ permutations\ of\ }abc )\right]-g_2\, 
{C_{ad}}^e{C_{be}}^f{C_{cf}}^d=0,
\ee
which is graphically represented in Fig.\ref{fig:eom}.
As shown in \cite{Sasakura:2005js}, the equation of motion for $g^{ab}$ is automatically satisfied due to the 
$GL(n,R)$ symmetry of the action, provided $g^{ab}$ is not singular and the equation of motion for $C_{abc}$
is satisfied.
 
\begin{figure}
\begin{center}
\includegraphics[scale=.5]{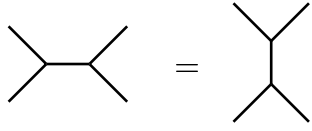}
\caption{The graphical representation of the associativity \eq{associativity}.}
\label{fig:associativity}
\end{center}
\end{figure}

An interesting case occurs when $g_1=g_2\neq 0,\ g_0\neq 0$. In this case, if the rank-three tensor satisfies 
the associativity condition \eq{associativity}, the second and third  
terms in the equation of motion 
\eq{eom} cancel with each other. This can be checked by using the graphical representation of the associativity 
condition in Fig.\ref{fig:associativity} to reconnect the diagrams in Fig.\ref{fig:eom}. Therefore, the presence
of the first term in \eq{eom} guarantees that the algebraic relation obtained from a classical solution 
is commutative {\it nonassociative}. 
This is important, because a commutative associative space with a positive (negative) definite $g^{ab}$ 
is trivial in the sense discussed in the following
paragraph. Therefore the case $g_1=g_2\neq 0,\ g_0\neq 0$ selects automatically non-trivial
commutative nonassociative fuzzy spaces as classical solutions.

Let me show the triviality of a commutative associative fuzzy
space which has a positive (negative) definite $g^{ab}$ and 
a finite number of functions\footnote{The usual continuum Euclidean space is a commutative associative space
with a positive definite $g^{ab}$, 
but is not trivial, because there are an infinite number of functions.}.
By using the $GL(n,R)$ transformation, a positive (negative) definite $g^{ab}$ can be diagonalized to
$g^{ab}=\delta_{ab}$ ($g^{ab}=-\delta_{ab}$). Then, 
because of the commutativity ${C_{ab}}^c={C_{ba}}^c$, the associativity \eq{associativity} implies
\be
N_a N_b=N_b N_a,
\ee
where $N_a$ is a symmetric matrix defined by ${(N_a)}_{bc} \equiv C_{abc}$. This mutual commutativity 
of $N_a$ implies that these matrices can be simultaneously diagonalized by an orthogonal matrix $M \in O(n,R)$,
which keeps $g^{ab}=\pm \delta_{ab}$ invariant,  
\be
{M_b}^{b'} {M_c}^{c'} {(N_a)}_{b'c'}=e(a,b)\, \delta_{bc},
\ee
where $e(a,b)$ denotes the diagonal elements.
Because of the commutativity ${C_{ab}}^c={C_{ba}}^c$, this $M$ will transform $C_{abc}$ to a totally diagonal form,
\be
C'_{abc}= {M_a}^{a'}{M_b}^{b'} {M_{c}}^{c'} C_{a'b'c'}=e(a)\, \delta_{ab}\, \delta_{bc}.
\ee
This shows that a commutative associative fuzzy space which has a positive (negative) definite $g^{ab}$ and 
a finite number of functions is just a collection of independent points.
Conversely, if $C_{abc}$ and $g^{ab}$ have such diagonal forms, the algebra is obviously commutative associative.

Let me finally comment on the satiability of the solutions. 
The equation of motion \eq{eom} just implies that the solutions are the stationary points of the action \eq{totals}, but not that they are 
the local minima.
The analysis of the fluctuations around the classical solutions is required to study the stability, and one cannot expect that all the 
solutions are locally stable. 
In the Euclidean case (when $g^{ab}$ of a classical solution is positive definite), however,  
there exists an obvious way to define an action 
in which the solutions to \eq{eom} are always stable. 
This is given by
\be
\label{sstable}
S_{stable}=g_{aa'}g_{bb'}g_{cc'} \frac{\partial S}{\partial C_{abc}}\frac{\partial S}{\partial C_{a'b'c'}}.
\ee
This action will generate various Euclidean classical solutions to \eq{eom} on equal footing, i.e. as its local minima, $S_{stable}=0$.
 
On the other hand, it does not seem obvious whether a solution must be a stable one to become interesting. 
One of the motivation behind in the present research is to discuss gravity in terms of fuzzy space. 
It is well known that the conformal mode in the Einstein-Hilbert action has a reversed kinetic term and therefore
the solutions cannot be stable in the usual sense. Moreover, when a Minkowski case is considered, 
the stationary points will be interesting in general. Therefore the preference of the choice of an action will
depend on the problems in mind. In this paper, the interest is only in the classical solutions to the equation of motion 
\eq{eom}, and the exact form of the action is not cared.

\section{An analytic example of a commutative nonassociative space}
\label{simpleexample}
In this section I will give an analytically treatable example of a commutative nonassociative fuzzy space. 
Although this space is not a solution to \eq{eom} and the number of functions on it is infinite, 
it has the properties very similar to the numerical solutions in the following section. 
Therefore this example will provide a good reference to understand the qualitative 
features of the numerical solutions.

The usual space is characterized by the following algebra of product,
\be
e^{ip^\mu x_\mu}e^{iq^\mu x_\mu}=e^{i(p^\mu+q^\mu) x_\mu},
\ee
where $e^{ip^\mu x_\mu}$ is a plane wave with momentum $p^\mu$ on a space coordinated by $x_\mu$.
I consider only Euclidean spaces.
In \cite{Sasai:2006ua}, the following commutative nonassociative deformation is 
considered\footnote{An apparent difference between \eq{deformprod} and the expression in \cite{Sasai:2006ua}
comes from the distinct normalizations of the functions and is essentially irrelevant.},
\be
\label{deformprod}
e^{ip^\mu x_\mu}*e^{iq^\mu x_\mu}=e^{-\alpha (p^2+q^2+(p+q)^2)} e^{i(p^\mu+q^\mu) x_\mu},
\ee
where $p^2=p^\mu p_\mu$, and $\alpha$ is a positive deformation parameter. 
To describe the deformed space,
the index of $C_{abc}$ and $g^{ab}$ of the tensor model 
is given by the  continuous momentum $p^\mu$, and their values are taken as 
\bea
\label{examplecg}
C_{p_1 p_2 p_3}&=&\delta^D(p_1+p_2+p_3)\, e^{-\alpha(p_1^2+p_2^2+p_3^2)}, \nonumber \\
g^{\,p_1p_2}&=&\delta^D(p_1+p_2),
\eea
where $D$ is the dimensions of the space, and the vector indices of momenta are suppressed for notational simplicity.
The measure of the momentum integration is defined by $\int d^Dp$.

In \cite{Sasakura:2005js},
the reality of the functions $f_a$ is assumed, which is the basis of the reality of $C_{abc}$ and $g^{ab}$. 
However the expression \eq{examplecg} is not in such a real coordinate, because $e^{ipx}$ is a complex function. 
An appropriate real coordinate can be obtained by considering instead the real functions, 
$\cos (px)= (e^{ipx}+e^{-ipx})/2$ and $\sin(px)=(e^{ipx}-e^{-ipx})/(2i)$. Then the indefinite $g^{ab}$ in
\eq{examplecg} is transformed to a positive definite one. For the computation in this section, 
however, the above complex coordinate is much more convenient, 
and can be used without confusion. On the other hand, the relation to such a real coordinate must be 
taken into account in the analysis of fluctuations around classical solutions, as done in \cite{Sasakura:2005gv}.
 
There is a simple useful tensor characterizing a fuzzy space, 
\be
\label{defofk}
K^a_b\equiv C^{acd}C_{bcd}.
\ee 
In the present case, putting \eq{examplecg} to \eq{defofk}, one obtains
\be
\label{ksimple}
K^{p_1}_{p_2}=\left( \frac{\pi}{4\alpha}\right)^{\frac{D}{2}}e^{-3\alpha p_1^2}\, \delta^D(p_1-p_2).
\ee
It is interesting to note that the factor depending on the momentum in \eq{ksimple} has the form of the heat kernel of the Laplacian. 

As discussed in \cite{Sasakura:2004dq,Sasakura:2005px}, 
the heat kernel expansion is a powerful method to obtain 
the low-momentum effective geometry of a fuzzy space. This method requires an operator 
on a fuzzy space which corresponds to the heat kernel of the Laplacian 
(or other elliptic differential operators). 
An indispensable feature of the Laplacian is the invariance 
under the general coordinate transformation. 
As for a fuzzy space, the properties of $K^a_b$ is invariant under the $GL(n,R)$ transformation, 
which is a fuzzy analog of the general coordinate transformation in general relativity.
Therefore, the operator $K^a_b$ will be the primary candidate to be used in applying this heat kernel method 
to derive the low-momentum effective geometry of a fuzzy space in general.

More concretely, the general procedure to obtain the effective geometric quantities of a fuzzy 
space can be described as follows. 
Suppose $K^a_b$ is diagonalizable with positive 
eigenvalues\footnote{This is true for a positive definite $g^{ab}$, 
which is the case studied in this paper. 
The definition \eq{defofk} shows that $K_{ab}$ is symmetric and (semi-)positive definite. 
Therefore, after diagonalizing $g^{ab}$ to $g^{ab}=\delta_{ab}$ by $GL(n,R)$ transformation, 
$K_{ab}$ can be transformed to a diagonal form with positive (or vanishing) diagonal elements 
by the remaining $O(n,R)$ symmetry which keeps $g^{ab}=\delta_{ab}$. 
The part with vanishing eigenvalues should be simply discarded in defining $K(t)$.}. 
Then let me define
an operator $K(t)^a_b$ depending on a parameter $t$ by
\be
K(t)^a_b= \sum_i \exp \left[t (\log k_i-\log k_{max})\right]\,(e_i)^a \, (\tilde e_i)_b, 
\label{defofkt}
\ee
where $k_i$ and $e_i$ are the eigenvalues and the eigenvectors of $K^a_b$, respectively, 
$k_{max}$ is the maximum of the eigenvalues,
and $\tilde e_i$ are the dual vectors of $e_i$, 
\be
(\tilde e_i)_a (e_j)^a=\delta_{ij}. 
\ee
The operator $K(t)$ will be assumed to give a fuzzy analogue of a heat kernel of an invariant operator in low-momentum.
Then the effective geometric quantities of a fuzzy space such as local curvatures will be obtained by investigating
the approximate asymptotic expansion of  ${\rm Tr}({\cal O}K(t))$ for appropriate 
insertion operators ${\cal O}$, as was discussed in \cite{Sasakura:2004dq,Sasakura:2005px}.

The philosophy behind in the above procedure is that an effective geometry should be 
defined by a probe. The probe here is a scalar field. Since $\{f_a\}$ span all the functions on a fuzzy space, 
a real scalar field will be described by a real rank-one tensor $\phi_a$. Therefore a candidate of 
an action of a non-self-interacting scalar field is given by  
\be
S_{scalar}=\phi^a (-K_{ab}+m_0^2 \, g_{ab}) \phi^b, 
\label{sscalar}
\ee
where $m_0$ is a constant. This action respects the invariance under the $GL(n,R)$ symmetry.
The dynamics of the scalar field from the action \eq{sscalar} will determine an 
effective geometry of a fuzzy space. 

Though the form of a scalar field action is constrained by the $GL(n,R)$ symmetry, 
there still remains 
large ambiguity, and this may harm the existence of a unique effective geometry.
In fact, the $K_{ab}$ in \eq{sscalar} should be replaced with its logarithm 
for a direct link to \eq{defofkt}.
Moreover $K_{ab}$ in \eq{sscalar} may be replaced with another rank-two symmetric tensor 
obtained from $C_{abc}$ and $g^{ab}$.
To justify the existence of a unique effective geometry,
the low-momentum limit of the various scalar field actions must have a universal character.
This is certainly true for the present example \eq{examplecg}.
Because of \eq{ksimple}, \eq{sscalar} approaches
the standard scalar field action in the low-momentum limit $p\rightarrow 0$.
Because of the Gaussian nature of \eq{examplecg}, it is obvious that another choice of the rank-two
tensor composed of $C_{abc}$ and $g^{ab}$ in \eq{examplecg} will also lead to a similar Gaussian form like \eq{ksimple}\footnote{
Because of the ambiguity of the construction of a scalar field action, one notices that the overall scaling of the metric cannot 
be determined. This is certainly natural, since there exist no intrinsic scales in the model. On the contrary, relative scale
will be meaningful.}.  In addition, it is numerically shown in the following section that the behavior of $K^a_b$ has 
a good fit with $const.\,e^{const.\, \Delta}$.
Although these evidences give only inconclusive support to the procedure
and the existence of a unique low-momentum effective geometry,
it can safely be insisted that at least some geometric properties can be incorporated in the 
commutative nonassociative algebra. This is in sharp contrast with the noncommutative geometry.
The matrix algebra itself does not contain any information on geometry. Various geometries results 
solely from the choice of Laplacian or matter field actions \cite{Balachandran:2005ew}.

\begin{figure}
\begin{center}
\includegraphics[scale=1]{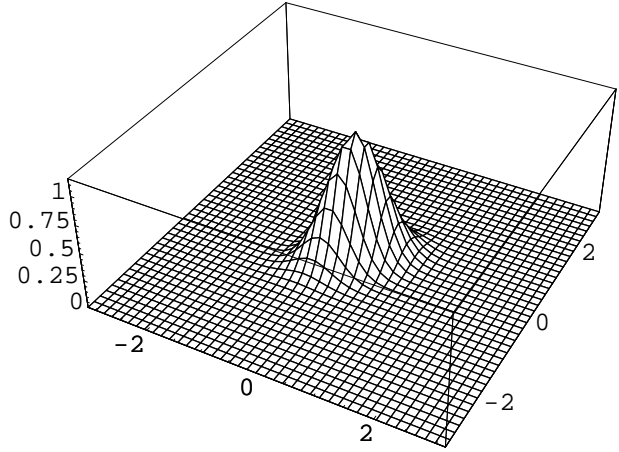}
\caption{$L(x,y)$ in \eq{lsimple} for $\alpha=\frac1{25}$ and $D=1$ is plotted.}
\label{fig:shapel}
\end{center}
\end{figure}

Now let me discuss another quantity which is not covariant or generally applicable, 
but is certainly useful to check the numerical results in the following section.
To avoid confusion, let me denote the functions on the fuzzy space by 
$f_p(=e^{ipx})$. Now let me define   
\be
d(x)\equiv \frac{1}{(2\pi)^D}\int d^Dp \, e^{-ipx} f_p,
\ee
where $e^{-ipx}$ is the usual $c$-number function and should not be 
confused with $f_{-p}$. This $d(x)$ 
is a fuzzy analog of the delta function with distribution on $x$.
The product of $d(x)$ and $d(y)$ can be written in the form,
\bea
d(x)*d(y)&=&\frac{1}{(2\pi)^{2D}} \int d^Dp_1 d^Dp_2 \, e^{-ip_1x-ip_2y}\,f_{p_1}*f_{p_2}   \\
&=&\frac{1}{(2\pi)^{2D}} \int d^Dp_1 d^Dp_2 d^Dq\, e^{-ip_1x-ip_2y}\,{C_{p_1p_2}}^{q} f_{q}   \\
&=& \int d^Dz \, L(x-z,y-z)\, d(z),
\eea
where 
\be
\label{fnl}
L(x,y)=\frac{1}{(2\pi)^{2D}}\int d^D p_1 d^D p_2 d^Dq  \, e^{-ip_1 x-ip_2 y}\, {C_{p_1p_2}}^{q}.
\ee
A straightforward computation for the present case \eq{examplecg} shows
\be
\label{lsimple}
L(x,y)= \frac{1}{\left(4\sqrt{3}\pi\alpha\right)^D} e^{-\frac{1}{12\alpha}\left( (x-y)^2+x^2+y^2 \right)}.
\ee 
As shown in Fig.\ref{fig:shapel},
for general $\alpha>0$, this two-parameter function has a peak at the origin $x=y=0$ and is distributed around it. 
The distribution is a little bit stretched in the direction $x=y$.
The profile becomes sharper for smaller $\alpha$, and 
in the associative limit $\alpha\rightarrow 0$, $L(x,y)$ approaches $\delta^D(x)\delta^D(y)$ as expected
from the product of two delta functions on an ordinary space.

As above, the existence of distribution of $L(x,y)$ around the origin can be regarded as a
sign of the fuzziness of a space. The profile will provide a quantitative measure of the fuzziness.
As the distribution becomes sharper, the fuzziness becomes smaller, and a fuzzy space approaches 
more an ordinary continuum space.
In the following section, $L(x,y)$ will be computed for the numerical solutions to check
how well the numerical solutions approximate ordinary continuum spaces. 

Let me finally discuss the general coordinate transformation. A general coordinate transformation is 
a transformation of the coordinate $x$ to a general invertible function of $x$,  
\be
\label{xpgx}
x'=g(x).
\ee
As discussed in \cite{Sasakura:2005js}, since the right-hand side of \eq{xpgx} can be generally expanded as 
a linear combination of the functions $f_a$, the analog of the general coordinate transformation on a fuzzy space
is given by the general linear transformation of $f_a$. 
On an ordinary space, the transformation of $e^{ipx}$ can be written in such a linear form as
\be
e^{ipx'}=
e^{ipg(x)}=\int d^Dq\, c(p,q)\,e^{iqx},
\ee
where
\be
\label{cpq}
c(p,q)=\frac1{(2\pi)^D}\int d^Dx\, e^{-iqx+ipg(x)}.
\ee
Therefore, the corresponding linear transformation for the present example is given by
\be
\label{fptrans}
f_p'=\int d^Dq\,c(p,q)\,f_q.
\ee
Then the product of the transformed functions is given by 
\bea
f_p'*f_q'&=&\int d^Dp'd^Dq'\, c(p,p')\,c(q,q')\,f_{p'}*f_{q'} \\
&=& \int d^Dr\, N(p,q;r)\, f_{r},
\eea
where
\bea
\label{npqr}
N(p,q;r)&=&\int d^Dp'd^Dq'\, c(p,p')\,c(q,q')\,{C_{p'q'}}^{r} \cr
&=& \int d^Dy\, c(p,y)\,c(q,r-y)\,e^{-\alpha\left(y^2+(r-y)^2+r^2\right)}.
\eea

From \eq{cpq} and \eq{npqr}, in the case of the ordinary flat space, $\alpha=0$, one can easily show that
\be
\label{npqreqcpqr}
N(p,q;r)=c(p+q,r),
\ee
and therefore, 
\be
f'_p*f'_q=f'_{p+q},
\ee
which implies that the product is invariant under the general coordinate transformation for $\alpha=0$.
On the contrary, $N(p,q;r)$ will depend not only on $p+q$ but generally both on $p$ and $q$ for $\alpha>0$, 
and the product cannot be invariant under the general coordinate transformation
except the obvious rotational and translational symmetry of \eq{deformprod}.

An essential distinction of the $\alpha>0$ case is the existence of the dimensionful parameter $\alpha$.
This dimensionful parameter characterizes the length scale of the fuzziness of the space. 
In another word, a kind of geometric structure is incorporated in the rank-three tensor $C_{abc}$ for
$\alpha>0$, while this is not for the ordinary space $\alpha=0$.
Therefore, as the metric tensor in general relativity is covariantly transformed, the rank-three tensor $C_{abc}$ 
should be transformed under the general coordinate transformation, which is nothing but the transformation \eq{ctrans}. 
This transformation \eq{ctrans} cancels that of the functions \eq{fptrans}, 
and the product between the functions becomes invariant under the general coordinate transformation.

At the classical solutions, 
the fuzzy analog of the general coordinate transformation symmetry will be broken except
some global remaining symmetries of a fuzzy space. 

\section{Numerical solutions}
\label{numericalsolutions}
In this section I will numerically obtain some solutions to the equation of motion \eq{eom}. By imposing the
expected global symmetries,
I obtain solutions corresponding to fuzzy flat torus and fuzzy spheres of various dimensions. To obtain the 
non-trivial solutions in the meaning discussed in Section \ref{themodel}, I only consider  the coupling
constants $g_0\neq 0,\ g_1=g_2\neq 0$. 
Since rescaling of $C_{abc}$ and the overall numerical factor of the action are irrelevant in the following 
analysis, I can put
\bea
\label{g012}
&&g_0=1,\cr
&&g_1=g_2=2,
\eea 
or $g_0=-1,g_1=g_2=2$ without loss of generality. 

In the latter case, however, the action has the form,
\be
S=\frac12 g^{aa'}{\rm Tr}\left( N_aN_{a'}\right) -\frac14 g^{aa'}g^{bb'}{\rm Tr}\left( [N_a,N_b][N_{a'},N_{b'}]
\right)
\ee 
with ${({N_a})_b}^c\equiv {C_{ab}}^c$.
If $g^{ab}$ is positive definite, this action is a sum of a positive definite term and a semi-positive 
definite one,
and cannot be stationary in the direction of rescaling $N_{a}$. 
In fact, only Euclidean spaces are considered in the following analysis,
and all the $g^{ab}$'s are  positive definite in the real coordinates mentioned in the preceding section.
Therefore the latter case does not have non-vanishing solutions.  
Thus it is enough to consider only the former case \eq{g012}.

Numerical solutions in this section have been obtained by taking $C_{abc}$ of ordinary spaces 
with an order-one normalization as an initial value, finding a local minimum of 
\be
V=\sum  \left(\,{\rm equation\ of\ motion}\,\right)^2,
\ee
where the sum is over all the equations of motion, 
and checking $V=0$ at the minimum within computational accuracy. 
Mathematica 5.0 on a desktop personal computer is used.  

In general relativity, the cosmological constant and the dimensions must be changed 
to obtain flat spaces and spheres of various dimensions as classical solutions. 
On the other hand, in this section, fuzzy flat torus and fuzzy spheres of various dimensions 
will be obtained as the classical solutions of the rank-three tensor model for the same coupling constants \eq{g012}. 
This implies that, in the rank-three tensor model, the cosmological constant 
and the dimensions are not fundamental, but should be regarded as dynamical quantities. 

\subsection{Solutions for commutative nonassociative fuzzy $S^1$ and $S^1\times S^1$}
As in the preceding section, the momentum index will be used. The momenta on a torus with unit radius take
integer values. They are cut-off to consider a finite number of dynamical variables. 
Thus the index of the model is assumed to be given by
\be
p=(n_1,n_2,\cdots,n_D), \ \ |n_i|\leq \Lambda,
\ee
where $n_i$ are integers, and $D$ is the dimensions of the torus. The cut-off $\Lambda$ may be different in each 
direction, but I consider only a common cut-off for simplicity.

A $D$-dimensional flat torus has the translational $U(1)^D$ symmetry. Under this symmetry, 
each function $f_p=e^{ipx}$ will be transformed by
\be
f'_p=e^{i n_i \theta^i} f_p,
\ee
where $\theta^i$ are the real parameters of the Lie group.
This symmetry is realized as $SO(2)^D$ symmetry in the 
real coordinate mentioned in the preceding section, and is the remaining symmetry of the 
breakdown of the $GL(n,R)$ symmetry.
Imposing the $U(1)^D$ symmetry on solutions, the non-vanishing components are restricted to the momentum 
conserving ones, $C_{p_1,p_2,-p_1-p_2}$ and $g^{p,-p}$. As discussed in \cite{Sasakura:2005js}, the $g^{ab}$ can
be fixed to a convenient value by virtue of the $GL(n,R)$ symmetry, and here let me take
\be
\label{gpmp}
g^{p,-p}=1.
\ee

A numerical solution to \eq{eom} with \eq{g012} and \eq{gpmp} for $\Lambda=5$ and $D=1$ is obtained as
\be
\label{flatsol}
\begin{array}{llll}
C_{-5, 0, 5}= 0.516893, & C_{-5, 1, 4}= 0.564918,& C_{-5, 2, 3}= 0.584882,& C_{-4,0, 4}=0.635328, \\
C_{-4, 1, 3}=0.665027,& C_{-4, 2, 2}=0.674038,& C_{-3, 0,3}=0.711723,& C_{-3, 1, 2}=0.728883, \\
C_{-2, 0, 2}=0.759477,& C_{-2, 1, 1}=0.76713,& C_{-1, 0, 1}=0.784838,& C_{0, 0, 0}=0.790784,
\end{array}
\ee
where the other momentum conserving components are obtained by the permutation symmetry \eq{constraintc} 
and another assumption on solutions, 
$C_{-p_1,-p_2,-p_2}=C_{p_1,p_2,p_3}$, which is a reflection symmetry of $S^1$. 

\begin{figure}
\begin{center}
\includegraphics[scale=1]{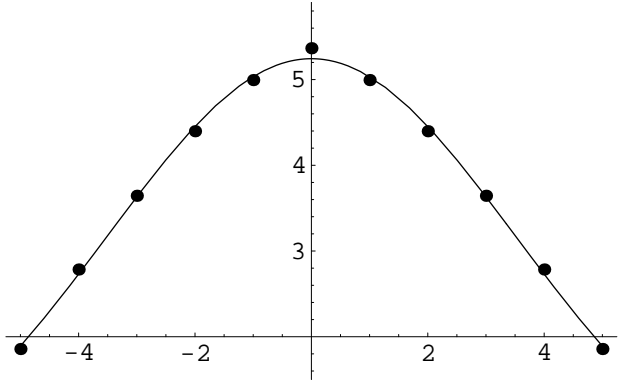}
\caption{$K_p^p$ is plotted for the fuzzy $S^1$ solution \eq{flatsol}. 
The solid line is $\exp(-0.041\, p^2+1.66)$.}
\label{fig:kflat}
\end{center}
\end{figure}

Using the solution \eq{flatsol}, $K_p^{p}$ defined in \eq{defofk} is plotted 
in Fig.\ref{fig:kflat}. The values can be fit well with $const.\,\exp(const. \Delta)$.

\begin{figure}
\begin{center}
\includegraphics[scale=1]{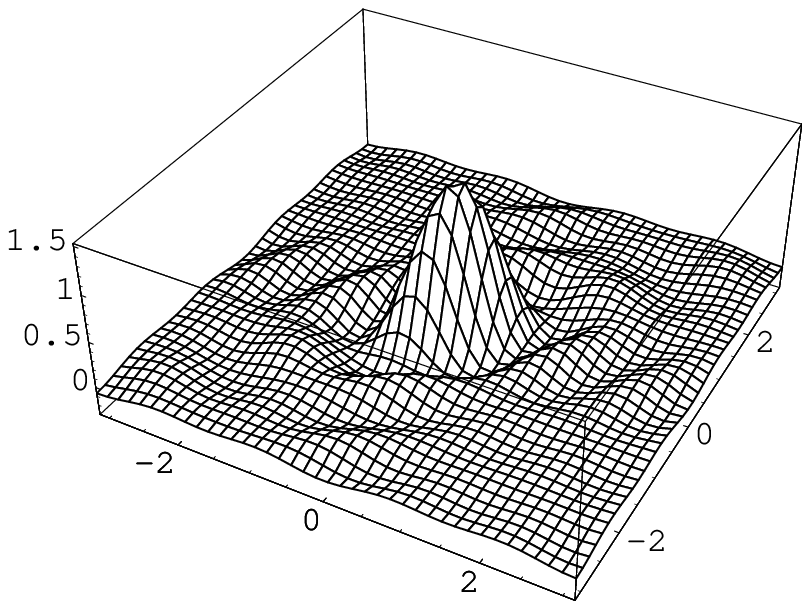}
\includegraphics[scale=1]{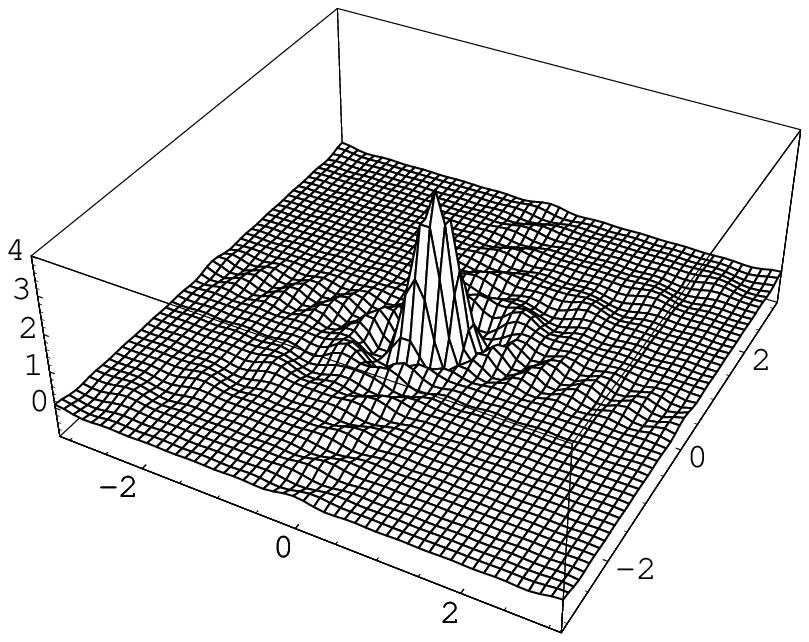}
\caption{$L(x,y)$ is plotted for the fuzzy $S^1$ with $\Lambda=5$ \eq{flatsol} in the left figure,
and that with $\Lambda=10$ in the right figure.
The peaks are well concentrated around the origin, 
similarly to that of the example fuzzy space in Fig.\ref{fig:shapel}.
As $\Lambda$ is increased, the peak becomes sharper, and the ordinary space will be obtained in the 
limit $\Lambda\rightarrow\infty$.}
\label{fig:lflat}
\end{center}
\end{figure}

The analog of the function $L(x,y)$ in \eq{fnl} for discrete momentum is defined by
\be
\label{fnldiscrete}
L(x,y)=\frac{1}{(2\pi)^{2D}}\sum_{p_1,p_2,q} \, e^{-ip_1 x-ip_2 y}\, {C_{p_1p_2}}^{q}.
\ee
This is plotted for the solution \eq{flatsol} in the left figure of Fig.\ref{fig:lflat}.
The profile is very similar to Fig.\ref{fig:shapel} of the example fuzzy space. One can check that,
as $\Lambda$ is increased, the profile becomes sharper as shown in the right figure of 
Fig.\ref{fig:lflat} for $\Lambda=10$, 
i.e. the ordinary space will be obtained in the limit $\Lambda\rightarrow\infty$.
   
\begin{figure}
\begin{center}
\includegraphics[scale=1]{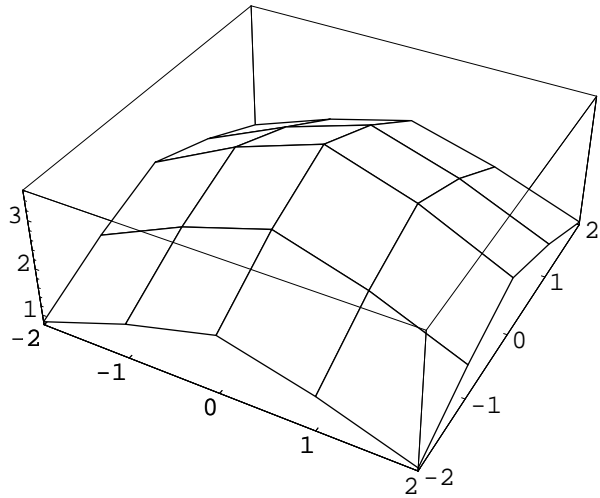}
\caption{$K^{(n_1,n_2)}_{(n_1,n_2)}$ is plotted for the fuzzy $S^1\times S^1$ solution \eq{flat2dimsol}. 
The vertices of the mesh are the values at the integer points $(n_1,n_2)$.}
\label{fig:kflat2dim}
\end{center}
\end{figure}

\begin{figure}
\begin{center}
\includegraphics[scale=1]{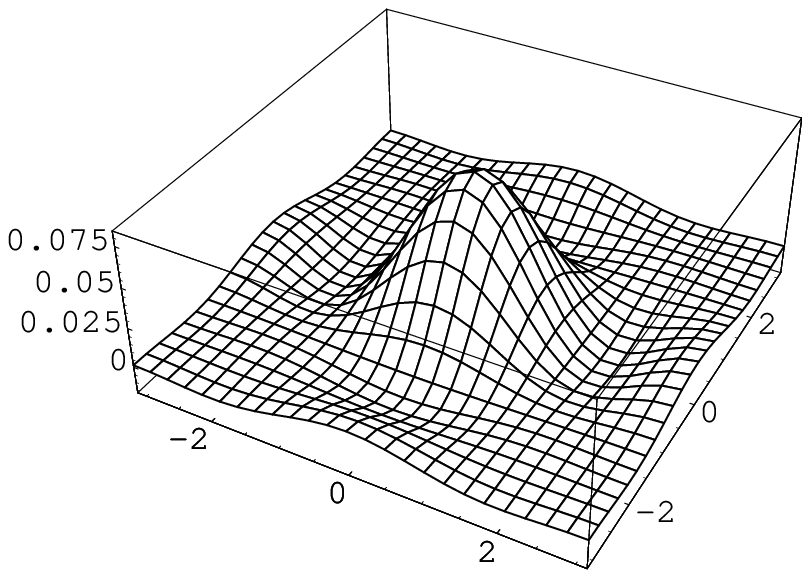}
\includegraphics[scale=1]{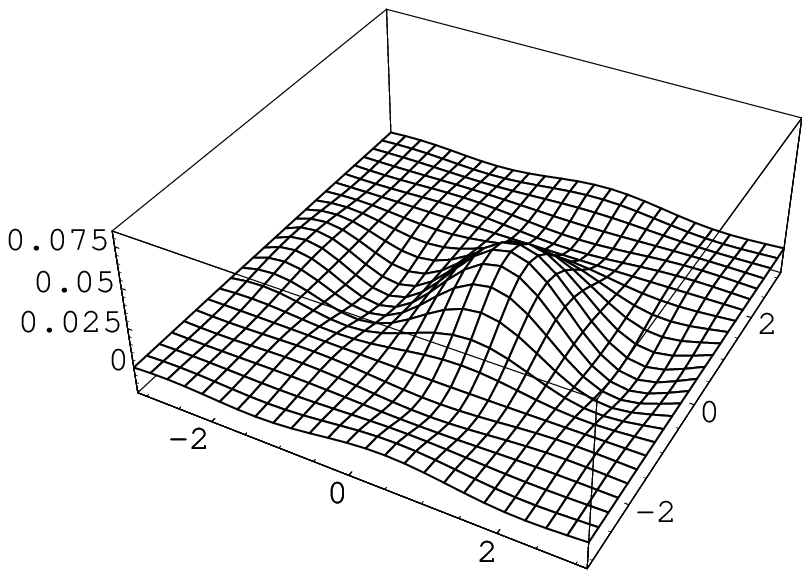}
\caption{$L((x_1,x_2),(0,0))$ (left) and $L((x_1,x_2),(1,0))$ (right) are plotted for the 
fuzzy $S^1\times S^1$ solution \eq{flat2dimsol}.
The qualitative behavior agrees with \eq{lsimple}.}
\label{fig:lflat2dim}
\end{center}
\end{figure}

The above kind of solution seems to exist also in higher dimensions. 
A numerical solution for $\Lambda=2$ and $D=2$ is obtained as 
\be
\label{flat2dimsol}
\begin{array}{lll}
C_{(2, -2), (0, 0), (-2, 2)}= 0.26524, & C_{(2, -2), (0, 1), (-2, 1)}=0.31018, & C_{(2, -2), (0, 2), (-2, 0)}=0.33439, \\ 
C_{(2, -2), (-1, 1), (-1, 1)}=0.31655,& C_{(2, -1), (0, 0), (-2, 1)}=0.35363,& C_{(2, -1), (0, 1), (-2, 0)}=0.37304,\\
C_{(2, -1), (0, 2), (-2, -1)}=0.36306,& C_{(2, -1), (-1, -1), (-1, 2)}=0.35707,& C_{(2, -1), (-1, 0), (-1, 1)}=0.37615,\\
C_{(2, 0), (0, 0), (-2, 0)}=0.38327, & C_{(2, 0), (0, 1), (-2, -1)}=0.37304, & C_{(2, 0), (-1, -1), (-1, 1)}=0.39036,\\ 
C_{(2, 0), (-1, 0), (-1, 0)}=0.39675,& C_{(2, 1), (0, 1), (-2, -2)}=0.31018, & C_{(1, -1), (0, 0), (-1, 1)}=0.41777, \\
C_{(1, -1), (0, 1), (-1, 0)}=0.42284,& C_{(1, 0), (1, 1), (-2, -1)}=0.37615,& C_{(1, 0), (0, 0), (-1, 0)}=0.43698,\\ 
C_{(0, 0), (0, 0), (0, 0)}=0.45174, &&
\end{array}
\ee
where all the other momentum conserving components are obtained by the symmetry \eq{constraintc} and another assumption
on solutions,
$C_{(n_1,n_1'),(n_2,n_2'),(n_3,n_3')}=C_{(-n_1,n_1'),(-n_2,n_2'),(-n_3,n_3')}=C_{(n_1',n_1),(n_2',n_2),(n_3',n_3)}$.
This assumption will be valid when the torus is a flat 2-torus $S^1\times S^1$, and the two $S^1$'s are equivalent
and symmetric under a reflection.
In Fig.\ref{fig:kflat2dim} and Fig.\ref{fig:lflat2dim},  $K^{(n_1,n_2)}_{(n_1,n_2)}$ and 
$L((x_1,x_2),(y_1,y_2))$ are plotted, respectively. 
The features of the solution again agrees well with those of the example fuzzy space in the preceding section. 
It has been checked that the shape of the peak becomes sharper as $\Lambda$ is increased.
Therefore the ordinary continuum space will be obtained in the $\Lambda\rightarrow\infty$ limit. 

In the ordinary coordinate description of spacetime, the number of the coordinates cannot be 
a dynamical quantity. In the tensor model, however, there exist flat torus solutions of various dimensions
for the same coupling constants \eq{g012}. 
Therefore the dimensions can be regarded as a dynamical quantity in the tensor model.

\subsection{Solutions for commutative nonassociative fuzzy $S^2$ and $S^3$}
In this section, I will show the existence of fuzzy spheres as classical solutions of the tensor model
for the same coupling constants \eq{g012}. This shows that the cosmological constant is also a dynamical
quantity in the tensor model.

A complete set of independent functions on ordinary $S^2$ is given by the set of spherical harmonic functions. 
The harmonic functions can be
classified by the index $(j,m)$, where $j$ is an integer spin of the $SO(3)$ rotation symmetry of $S^2$ and 
$m$ is a spin component in a direction, $m=-j,-j+1,\cdots, j$.
Therefore, in this subsection, the index of the tensor model is assumed to be
\be
a=(j,m),\ \ \  |m|\leq j \leq \Lambda,
\ee
where $j,m$ are integers and $\Lambda$ is a cut-off.

The rank-three tensor of an $SO(3)$ invariant solution can be assumed to have a form,
\be
\label{defcjm}
C_{(j_1,m_1),(j_2,m_2),(j_3,m_3)}=A_{j_1,j_2,j_3} \,
\left(
\begin{array}{ccc}
j_1 & j_2 & j_3 \\
m_1 & m_2 & m_3 
\end{array}
\right),
\ee
where $A_{j_1,j_2,j_3}$ is real, its indices must satisfy the triangle inequality $|j_2-j_3|\leq j_1 \leq j_2+j_3$
for non-vanishing components, 
and $(:::)$ is the $3j$-symbol \cite{Messiah,Varshalovich:1988ye}.
The ordinary spherical harmonics have the product relation such that 
$A_{j_1,j_2,j_3}$ is non-vanishing only when $j_1+j_2+j_3$ is an even integer. 
Therefore I also assume 
\be
A_{j_1,j_2,j_3}=0\ {\rm for} \ j_1+j_2+j_3={\rm odd}.
\ee
Then the permutation symmetry \eq{constraintc} and the permutation property of the $3j$-symbol imposes
\be
A_{j_1,j_2,j_3}=A_{j_{\sigma(1)},j_{\sigma(2)},j_{\sigma(3)}}
\ee 
for any permutation $\sigma$.
The rank-two tensor is taken as 
\be
\label{defgjm}
g^{(j_1,m_1), (j_2,m_2)}=\delta_{j_1j_2}\delta_{m_1+m_2,0}\, (-1)^{j_1+m_1},
\ee
which is invariant under the $SO(3)$ symmetry.
As shown explicitly in \cite{Sasakura:2005gv}, the expressions \eq{defcjm} and \eq{defgjm} in momentum coordinate
can be transformed to real valued components in a real coordinate.

By using some properties of $3j$-symbols \cite{Messiah,Varshalovich:1988ye} such as 
\be
\sum_{m_1,m_2} 
\left(
\begin{array}{ccc}
j_1 & j_2 & j_3 \\
m_1 & m_2 & m_3 
\end{array}
\right)
\left(
\begin{array}{ccc}
j_1 & j_2 & j'_3 \\
m_1 & m_2 & m'_3 
\end{array}
\right)
=
\frac{1}{2j_3+1}\delta_{j_3j_3'} \delta_{m_3m_3'}
\ee
and 
\bea
\sum_{M_i}
(-1)^{\sum_i J_i+M_i}  
\left(
\begin{array}{ccc}
J_1 & J_2 & j_3 \\
M_1 & -M_2 & m_3 
\end{array}
\right)
\left(
\begin{array}{ccc}
J_2 & J_3 & j_1 \\
M_2 & -M_3 & m_1 
\end{array}
\right)
\left(
\begin{array}{ccc}
J_3 & J_1 & j_2 \\
M_3 & -M_1 & m_2 
\end{array}
\right) \cr
=\left\{ 
\begin{array}{ccc}
j_1 & j_2 & j_3 \\
J_1 & J_2 & J_3 
\end{array}
\right\},
\eea
where \{:::\} is the $6j$-symbol,
the equation of motion \eq{eom} can be rewritten as 
\bea
\label{eombya}
&&-g_0 A_{j_1,j_2,j_3} + \frac{g_1}{3} \left[ \frac{1}{2j_1+1} A_{j_1,j_2,j_3} \sum_{j_4,j_5} A_{j_1,j_4,j_5}^2 +({\rm cyclic\ permutations\ of}\ j_1j_2j_3)\right] \cr
&&-g_2 \sum_{j_4,j_5,j_6} 
\left\{ 
\begin{array}{ccc}
j_1 & j_2 & j_3 \\
j_4 & j_5 & j_6 
\end{array}
\right\}
A_{j_1,j_5,j_6} A_{j_2,j_4,j_6} A_{j_3,j_4,j_5}=0,
\eea
where the values of the couplings are given by \eq{g012}. 

A numerical solution to \eq{eombya} for $\Lambda=5$ is obtained as 
\be
\label{sols2}
\begin{array}{llll}
A_{0, 0, 0}=0.375888, & A_{0, 1, 1}=0.642392, & A_ {1, 1, 2}=0.80089, & A_{0, 2, 2}=0.893485, \\
A_{2, 2, 2}=0.92406, & A_{1, 2, 3}=0.865991, & A_{0, 3, 3}=0.887106,& A_{2, 3, 3}=1.03293, \\
A_{2, 2, 4}=1.11042, & A_{1, 3, 4}=1.08588, & A_{3, 3, 4}=0.947859, & A_{0, 4, 4}=1.15568, \\
A_{2, 4, 4}=1.03207, & A_{4, 4, 4}=1.17188, & A_{2, 3, 5}=1.08615, & A_{1, 4, 5}=1.0445, \\
A_{3, 4, 5}=1.07613, & A_{0, 5, 5}=1.09768, & A_{2, 5, 5}=1.09539, & A_{4, 5, 5}=1.10454.
\end{array}
\ee

\begin{figure}
\begin{center}
\includegraphics[scale=1]{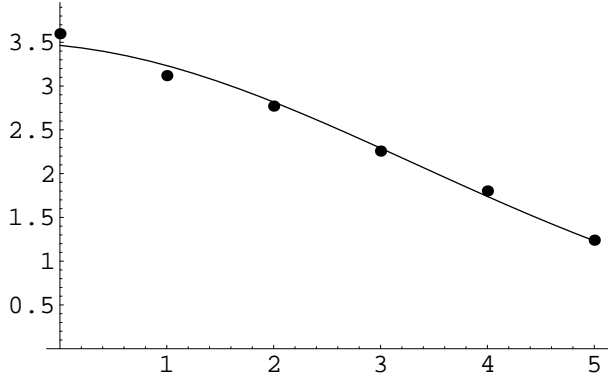}
\caption{$K^{(j,m)}_{(j,m)}$ is plotted for the fuzzy $S^2$ solution \eq{sols2}. The horizontal axis is $j$. The solid line is $\exp [ -0.035\, j(j+1)+1.24]$ }
\label{fig:props2}
\end{center}
\end{figure}

Using the properties of the $3j$-symbol, the quantity \eq{defofk} can be rewritten as 
\bea
\label{ks2witha}
K^{(j,m)}_{(j',m')}&=&\sum_{j_1,j_2,m_1,m_2} C^{(j,m),(j_1,m_1),(j_2,m_2)}C_{(j',m'),(j_1,m_1),(j_2,m_2)} \cr
&=& \delta^j_{j'}\delta^m_{m'} \frac1{2j+1} \sum_{j_1,j_2} A_{j,j_1,j_2}^2,
\eea
which basically depends only on $j$ due to the $SO(3)$ symmetry. The $K^{(j,m)}_{(j,m)}$ for the solution
\eq{sols2} are plotted in Fig.\ref{fig:props2}.
The values can be fit well by $const.\,\exp(const. \Delta)$.

In the following, let me compare the product relation obtained from the solution \eq{sols2} 
and that of the spherical harmonics.  
One needs to take care of the phase choice of the spherical harmonics. 
The choice compatible with \eq{defgjm} is such that  
the spherical harmonics 
satisfy\footnote{The spherical harmonics with this phase convention are referred to as modified spherical harmonics in 
\S 5.1.5 of \cite{Varshalovich:1988ye}.}
\be
\tilde Y_{j,m}^*(\Omega)=(-1)^{j+m} \tilde Y_{j,m}(\Omega)
\ee
under the complex conjugation.
For these spherical harmonics, the rank-three tensor defining the product is given by
\bea
&&\tilde C_{(j_1,m_1),(j_2,m_2),(j_3,m_3)}=
\int d\Omega\, \tilde Y_{j_1,m_1}(\Omega)\tilde Y_{j_2,m_2}(\Omega)\tilde Y_{j_3,m_3}(\Omega) \cr
&&\ \ \ \ \ \ \ \ \ \ \ = (-1)^{(j_1+j_2+j_3)/2}
\left[\frac{\prod_i (2j_i+1)}{4\pi}\right]^{\frac12}
\left(
\begin{array}{ccc}
j_1 & j_2 & j_3 \\
0&0&0
\end{array}
\right)
\left(
\begin{array}{ccc}
j_1 & j_2 & j_3 \\
m_1&m_2&m_3
\end{array}
\right).
\eea
Therefore the quantity of the ordinary $S^2$ corresponding to $A_{j_1,j_2,j_3}$ is given by
\be
\label{ats2}
\tilde A_{j_1,j_2,j_3} = (-1)^{(j_1+j_2+j_3)/2}
\left[\frac{\prod_i (2j_i+1)}{4\pi}\right]^{\frac12}
\left(
\begin{array}{ccc}
j_1 & j_2 & j_3 \\
0&0&0
\end{array}
\right).
\ee

\begin{figure}
\begin{center}
\includegraphics[scale=1]{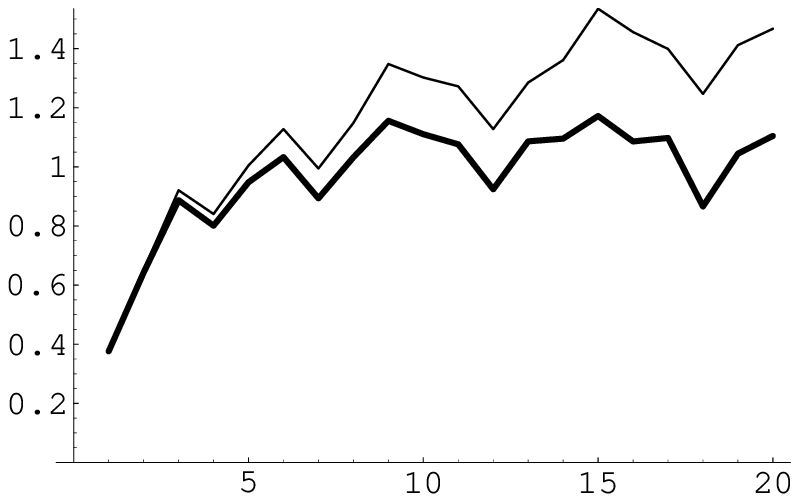}
\includegraphics[scale=1]{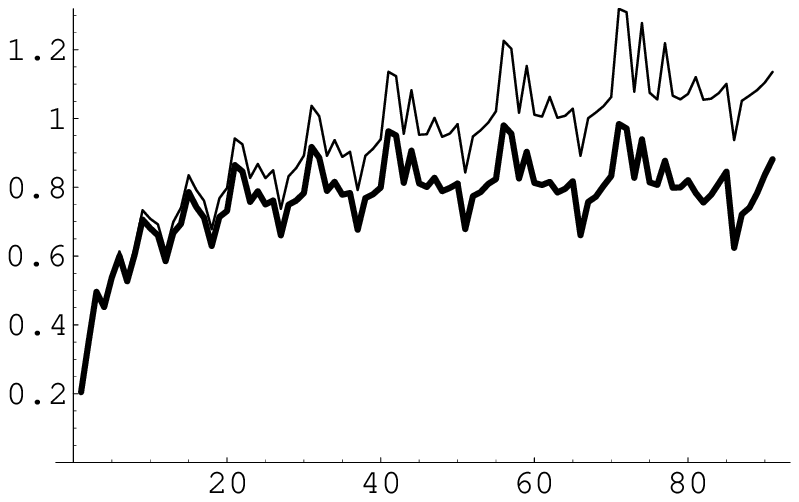}
\caption{In the left figure, $A_{j_1,j_2,j_3}$ of the fuzzy $S^2$ solution with $\Lambda=5$ in \eq{sols2} 
(bold line) and $\tilde A_{j_1,j_2,j_3} A_{0,0,0}/ \tilde A_{0,0,0}$ of the ordinary $S^2$ in \eq{ats2} 
(thin line) are plotted.
The values are ordered in the same order as \eq{sols2}, i.e. smaller to larger spins. 
The discrete points are joined by lines for clear view.
The solution has a tendency to take smaller values at larger spins.
In the right figure, the same for a solution with $\Lambda=10$. 
Comparing the regions below 20 in each figure, it can be observed that 
a better approximation to the ordinary $S^2$ is obtained 
for $\Lambda=10$ than $\Lambda=5$.}
\label{fig:aats2}
\end{center}
\end{figure}

In the left figure of Fig.\ref{fig:aats2}, 
the solution $A_{j_1,j_2,j_3}$ in \eq{sols2} and $\tilde A_{j_1,j_2,j_3}$ of the ordinary sphere in \eq{ats2} 
are compared. 
A similar diagram is shown for the solution with $\Lambda=10$ in the right figure of Fig.\ref{fig:aats2}. 
One can see that the solutions approximate well the ordinary product between spherical harmonics.
The approximation becomes better for $\Lambda=10$ than $\Lambda=5$.
Therefore, the solution can be identified with a fuzzy $S^2$, and 
the ordinary continuum $S^2$ will be obtained in the limit $\Lambda\rightarrow \infty$.

\begin{figure}
\begin{center}
\includegraphics[scale=1]{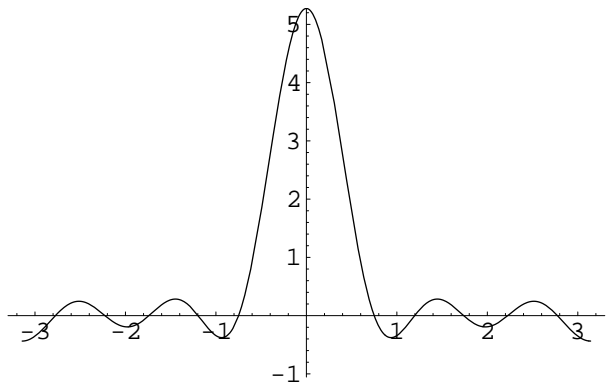}
\caption{$L(\Omega_0,\Omega)$ for the fuzzy $S^2$ solution \eq{sols2} is plotted. 
The horizontal axis is the angle between $\Omega_0$
and $\Omega$. The values are well concentrated around the origin, that is consistent with the product between
two fuzzy delta functions at $\Omega_0$.}   
\label{fig:ls2}
\end{center}
\end{figure}

An analogous quantity for a sphere corresponding to \eq{fnl} can be defined by
\be
L(\Omega,\Omega')\equiv \sum_{j_i,m_i} C_{(j_1,-m_1),(j_2,-m_2),(j_3,-m_3)}\tilde Y_{j_1,m_1}(\Omega_0) 
\tilde Y_{j_2,m_2}(\Omega) \tilde Y_{j_3,m_3}(\Omega'),
\ee
where $\Omega_0$ denotes a reference point on $S^2$, and one of the delta functions is located there. 
In Fig.\ref{fig:ls2}, $L(\Omega_0,\Omega)$ for the $\Lambda=5$ solution \eq{sols2} is plotted. 
The profile is consistent with the fuzzy product between two delta functions at $\Omega_0$.

Let me finally discuss a fuzzy $S^3$ solution. 
The spherical harmonics on $S^3$ are given by the complete set of representation matrices of $SU(2)$, 
$D_{mn}^j(g)$, $g\in SU(2)$ with $j=0,\frac12,1,\cdots$ \cite{Dowker:2004nh}. These Wigner $D$-functions satisfy
\cite{Varshalovich:1988ye} 
\be
\label{intddd}
\int dg\, D^{j_1}_{m_1n_1}(g)  D^{j_2}_{m_2n_2}(g)  D^{j_3}_{m_3n_3}(g)
=\left( 
\begin{array}{ccc}
j_1 & j_2 & j_3 \\
m_1 & m_2 & m_3
\end{array}
\right)  
\left( 
\begin{array}{ccc}
j_1 & j_2 & j_3 \\
n_1 & n_2 & n_3
\end{array}
\right),
\ee  
where the structure of the combination of two $3j$-symbols comes from the invariance under the two $SU(2)$'s of 
$SO(4) \sim  SU(2) \times SU(2) /Z_2$. Therefore I assume the solution to have the 
form\footnote{
The spin $j_i$ must be common in the two $3j$-symbols for a solution to represent an $S^3$.
If they were allowed to be different, a solution would represent $S^2\times S^2$ instead of $S^3$.},
\be
C_{(j_1,m_1,n_1),(j_2,m_2,n_2),(j_3,m_3,n_3)}=B_{j_1,j_2,j_3} 
\left( 
\begin{array}{ccc}
j_1 & j_2 & j_3 \\
m_1 & m_2 & m_3
\end{array}
\right)  
\left( 
\begin{array}{ccc}
j_1 & j_2 & j_3 \\
n_1 & n_2 & n_3
\end{array}
\right),
\ee
where $B_{j_1,j_2,j_3}$ is assumed to be real and symmetric under any permutation of 
the indices.   

The rank-two tensor is taken as
\be
g^{(j_1,m_1,n_1),(j_2,m_2,n_2)}=\delta_{j_1,j_2}\delta_{m_1+m_2,0} \delta_{n_1+n_2,0} 
(-1)^{m_1-n_1},
\ee
which is just obtained by multiplying \eq{defgjm} and its inverse. This choice is consistent with
the property of the Wigner $D$-functions, 
\be
\label{dnormalization}
\int dg\, D^{j_1}_{m_1n_1}(g) D^{j_2}_{m_2n_2} (g) =\frac{\delta_{j_1,j_2}\delta_{m_1+m_2,0}
\delta_{n_1+n_2,0}\,(-1)^{m_1-n_1}}{2j_1+1} 
\ee
up to the factor $\frac{1}{2j_1+1}$, which will be considered in due course.
Because of these double-structures, the equation of motion in 
$B_{j_1,j_2,j_3}$ can be just obtained by taking the squares of the coefficients in \eq{eombya},
\bea
\label{eombyat}
&&-g_0 B_{j_1,j_2,j_3} + \frac{g_1}{3} \left[ \frac{1}{(2j_1+1)^2} B_{j_1,j_2,j_3} \sum_{j_4,j_5} 
B_{j_1,j_4,j_5}^2 +({\rm cyclic\ permutations\ of}\ j_1j_2j_3)\right] \cr
&&-g_2 \sum_{j_4,j_5,j_6} 
\left\{ 
\begin{array}{ccc}
j_1 & j_2 & j_3 \\
j_4 & j_5 & j_6 
\end{array}
\right\}^2
B_{j_1,j_5,j_6} B_{j_2,j_4,j_6} B_{j_3,j_4,j_5}=0.
\eea

The spin will be cut-off by $j\leq \Lambda$, 
and a numerical solution for $\Lambda=3$ is obtained as 
\be
\label{sols3}
\begin{array}{llll}
B_{0, 0, 0}      = 0.179229, &
B_{0, 1/2, 1/2}  = 0.352866, &
B_{1/2, 1/2, 1}  = 0.516800, &
B_{0, 1, 1}      = 0.661060, \\
B_{1, 1, 1}      = 0.784606, &
B_{1/2, 1, 3/2}  = 0.860387, &
B_{0, 3/2, 3/2}  = 0.899459, &
B_{1, 3/2, 3/2}  = 0.600820, \\
B_{1, 1, 2}      = 0.819278, &
B_{1/2, 3/2, 2}  = 1.018466, &
B_{3/2, 3/2, 2}  = 1.153431, &
B_{0, 2, 2}      = 1.252915, \\
B_{1, 2, 2}      = 0.889080, &
B_{2, 2, 2}      = 1.078355, &
B_{1, 3/2, 5/2}  = 1.150590, &
B_{1/2, 2, 5/2}  = 1.275880, \\
B_{3/2, 2, 5/2}  = 1.393015, &
B_{0, 5/2, 5/2}  = 1.419768, &
B_{1, 5/2, 5/2}  = 1.559359, &
B_{2, 5/2, 5/2}  = 1.683407, \\
B_{3/2, 3/2, 3}  = 1.428888, &
B_{1, 2, 3}      = 1.497101, &
B_{2, 2, 3}      = 1.687474, &
B_{1/2, 5/2, 3}  = 1.851444, \\
B_{3/2, 5/2, 3}  = 1.738983, &
B_{5/2, 5/2, 3}  = 1.964490, &
B_{0, 3, 3}      = 2.013377, &
B_{1, 3, 3}      = 2.241499, \\
B_{2, 3, 3}      = 2.368063, &
B_{3, 3, 3}      = 2.867651. & \ &
\end{array}
\ee
\begin{figure}
\begin{center}
\includegraphics[scale=1]{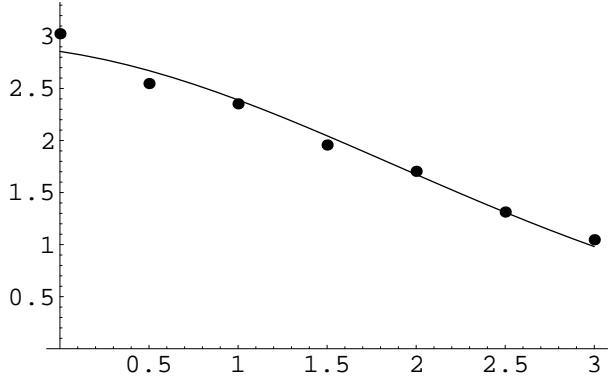}
\caption{$K^{(j,m,n)}_{(j,m,n)}$ for fuzzy $S^3$ is plotted. The horizontal axis is $j$.
The solid line is $\exp [ -0.089\, j(j+1)+1.05]$}
\label{fig:ks3}
\end{center}
\end{figure}

By taking the square of the coefficient in \eq{ks2witha}, the $K^a_b$ is obtained as 
\be
\label{ks3witha}
K^{(j,m,n)}_{(j',m',n')}
= \delta^j_{j'}\delta^m_{m'} \delta^n_{n'} \frac1{(2j+1)^2} \sum_{j_1,j_2} B_{j,j_1,j_2}^2,
\ee
which depends essentially only on $j$. The values for the solution \eq{sols3} 
are plotted in Fig.\ref{fig:ks3}. The values can be fit well with $const.\,\exp(const. \Delta)$.

\begin{figure}
\begin{center}
\includegraphics[scale=1]{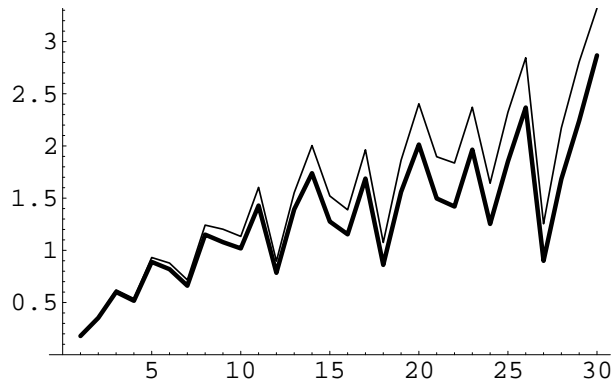}
\caption{The fuzzy $S^3$ solution, $B_{j_1,j_2,j_3}$ in \eq{sols3}, (bold) and 
the values of the ordinary $S^3$, $\tilde B_{j_1,j_2,j_3} B_{0,0,0}/ \tilde B_{0,0,0}$ in \eq{ords3},
(thin) are compared. The values are ordered in the same order as \eq{sols3}, i.e. smaller to larger spins. 
The discrete points are joined by lines for clear view. The solution has a tendency to take smaller values at larger spins.}
\label{fig:s3comp}
\end{center}
\end{figure}

After including the factor appearing in \eq{dnormalization} into the normalization of the spherical harmonics, 
it can be shown from \eq{intddd} that the quantity of the ordinary $S^3$ corresponding to $B_{j_1j_2j_3}$ is given by
\be
\label{ords3}
\tilde B_{j_1,j_2,j_3}= \left[ \prod_i (2j_i+1) \right]^{\frac12}.
\ee
In Fig.\ref{fig:s3comp}, this $\tilde B_{j_1,j_2,j_3}$ is compared with $B_{j_1,j_2,j_3}$ of the solution \eq{sols3}. 
A good agreement is observed, and the solution \eq{sols3} can be identified with a fuzzy $S^3$.

\section{Summary and discussions} 
In this paper, the proposal in \cite{Sasakura:2005js} that the rank-three tensor model 
can be used as theory of dynamical fuzzy spaces is applied to dynamical generation of 
commutative nonassociative fuzzy spaces.
It is numerically shown that fuzzy flat torus and fuzzy spheres of various dimensions 
are classical solutions of the rank-three tensor model. 
Therefore, in the tensor model, the cosmological constant
and the dimensions are not fundamental but can be regarded as dynamically generated quantities.
It is also observed that,
as the number of the degrees of freedom increases, the fuzziness becomes smaller, and 
the ordinary spaces will be recovered in the infinite limit. 
 
In the tensor model, $GL(n,R)$ is a fuzzy analog of the general coordinate transformation symmetry in
general relativity. There is an old idea 
that gravity may be regarded as Nambu-Goldstone fields of a spontaneous broken local translational 
invariance \cite{Borisov:1974bn,Boulanger:2006tg} 
in a similar manner as the Abelian gauge theory \cite{Ferrari:1971at,Brandt:1974jw}.
Since the $GL(n,R)$ symmetry is broken at classical solutions except some global symmetries,
the rank-three tensor model is a finite concrete model which may realize the idea. 
However, the applicability of the idea to the tensor model is not clear at present, since the discussions 
have been for ordinary continuous spacetime.
If it is shown to be applicable, the tensor model can be an interesting candidate for quantum gravity.

There was a serious unsatisfactory property in the solution of a {\it noncommutative} two-sphere 
in the tensor model  
found in the previous paper \cite{Sasakura:2005js}. 
In the discussions of the interpretation of $GL(n,R)$ symmetry as a fuzzy analog of 
the general coordinate transformation \cite{Sasakura:2005js}, it is essential that 
the set of functions $\{f_a\}$ span all the scalar functions on a fuzzy space. 
However, in the previous {\it noncommutative} solution, the scalar field had a double-index structure
$\phi_{ab}$, and could not be identified with $\{f_a\}$. 
On the contrary, the solutions found in this paper are {\it commutative nonassociative} deformations
of the algebraic relations of the functions on the ordinary spaces with truncation, 
and the $GL(n,R)$ transformation can be naturally identified with a fuzzy analog of 
the general coordinate transformation.

The interpretation of the rank-three tensor model as a model of fuzzy spaces has some advantages over the 
original interpretation of tensor models \cite{Ambjorn:1990ge,Sasakura:1990fs,Godfrey:1990dt,Boulatov:1992vp,Ooguri:1992eb,DePietri:1999bx,DePietri:2000ii,
Freidel:2005cg} that Feynman diagrams correspond to dual diagrams of simplicial complexes. 
In the original interpretation, the rank of tensor is related to the dimensions and must be 
changed to discuss quantum gravity in different dimensions. On the contrary, 
in the present interpretation, the rank is kept three, but various dimensions are dynamically induced.
In addition, in the original interpretation, a kind of topological expansion is required to obtain physical results 
from tensor models. However, such an expansion has not been found for tensor models. On the other hand, 
in the present interpretation, the physical meaning is definite. Moreover
the quantum fluctuations around a classical solution are controllable in principle by
taking a semi-classical limit, and several quantities will be computable.

There is a worry about the locality of the model. The behavior of $L(x,y)$ of the classical solutions shows
that the peak is well localized around the origin, i.e. the classical
solutions are local. However, the fluctuations around the classical solutions obviously contain various 
non-local modes. 
It seems inevitable that, since there exists no metric at the beginning, 
a pregeometric system like the tensor model contains 
modes which are non-local in the sense of the background of classical solutions.
Therefore the right question would be whether such non-local fluctuations 
lead to serious contradictions to the principles in physics. 
There are a number of ways to avoid contradictions like 
that non-local modes are very heavy or that interactions between local and non-local modes 
are negligibly small. 
A systematic analysis of fluctuation modes around classical solutions is required as a future work. 

\begin{figure}
\begin{center}
\includegraphics[scale=.7]{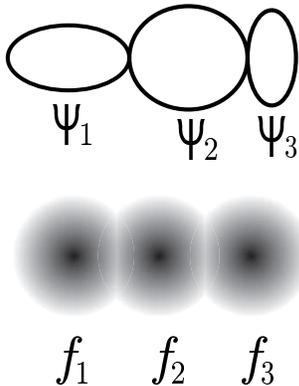}
\caption{The upper figure shows an example in closed string field theory 
for the nonassociativity, $(\Psi_2*\Psi_1)*\Psi_3\neq \Psi_2*(\Psi_1*\Psi_3)$, 
since $\Psi_1*\Psi_3=0$. Similar nonassociativity seems to hold for a fuzzy space in the lower figure.}   
\label{fig:nonassociative}
\end{center}
\end{figure}

The notion of nonassociative space is rather new, compared with noncommutative space. 
Noncommutativity is natural from the viewpoint of quantization, but 
nonassociativity is rather exceptional in physical systems. 
However, as explained briefly in the Introduction, nonassociativity can be also a physically sensible
structure of spacetime.  
In fact, as shown in Fig.\ref{fig:nonassociative}, the two figures for closed string field theory 
and a fuzzy space, respectively, have some similarities, and the intuitive argument on closed string field theory in
\cite{Witten:1985cc} seems to also hold for a fuzzy space. 
As discussed in Section \ref{simpleexample}, an interesting aspect of a nonassociative space 
is that a kind of metric structure is incorporated in the definition.
This would be the most peculiar property distinct from a noncommutative space which seems to be associated with 
a kind of symplectic structure.

The present construction of the tensor model is not unique. 
Although the action \eq{totals} is among the simplest, there is no reason to discard the higher order interactions.  
Another ambiguity is the choice of the action \eq{sscalar} of a scalar field on a fuzzy space. 
As discussed in Section \ref{simpleexample}, 
this ambiguity may harm the existence of a unique low-momentum effective geometry
defined by the probe of a scalar field. 
It might turn out to be true that 
the low-momentum dynamics does not depend on the details of the actions, as the universality in lattice theory.
However, although this kind of universality has a chance to hold near some classical solutions, it is hard to believe that
the full quantum properties such as transition probabilities between classical solutions do not depend on
the details.
Therefore, even if the tensor model has a chance to provide an interesting model of quantum gravity by itself, 
it seems very important to find its principle in relation with
other systematic approaches of quantum gravity or string theory. 
 
\vspace{.5cm}
\section*{Acknowledgments}
The author was supported in part by the Grant-in-Aid for Scientific Research No.13135213, No.16540244 and No.18340061
from the Ministry of Education, Science, Sports and Culture of Japan.

\vspace{1cm}
\noindent
{\bf Note added in proof: } I was informed that systematic construction of fuzzy spheres in diverse dimensions using quantized Nambu 
brackets, which is distinct from the formulation of this paper, was proposed, and their features were discussed 
in a series of papers \cite{jabbari}.

\end{document}